\documentclass[acmsmall]{acmart}

\usepackage{amsmath}
\usepackage{amssymb}
\usepackage{rotating}
\usepackage{booktabs}
\usepackage{listings}

\usepackage{algpseudocode}

\usepackage{algorithmicx}

\usepackage[ruled]{algorithm2e} % For algorithms

\SetAlFnt{\small}
\SetAlCapFnt{\small}
\SetAlCapNameFnt{\small}
\SetAlCapHSkip{0pt}
\IncMargin{-\parindent}
\AtBeginDocument{%
  \providecommand\BibTeX{{%
    \normalfont B\kern-0.5em{\scshape i\kern-0.25em b}\kern-0.8em\TeX}}}

\setcopyright{acmcopyright}
\copyrightyear{2020}
\acmYear{2020}
\acmDOI{}

\acmJournal{TOIT}
\acmVolume{1}
\acmNumber{1}
\acmArticle{1}
\acmMonth{1}

\begin{document}

\title{An Incentive-Based Mechanism for Volunteer Computing using Blockchain}

\author{Ismaeel Al Ridhawi}
\email{i.alridhawi@kcst.edu.kw}
%\email{i.alridhawi@kcst.edu.kw}
\orcid{0000-0001-5822-2763}
\affiliation{%
	\institution{Kuwait College of Science and Technology}
	%\streetaddress{Doha District}
	%\postcode{K1N6N5}
	\country{Kuwait}
}

\author{Moayad Aloqaily}
\email{maloqaily@ieee.org}
\orcid{0000-0003-2443-7234}
\affiliation{%
	\institution{Al Ain University}
	\country{UAE}
}

\author{Yaser Jararweh}
\email{YiJararweh@just.edu.jo}
\orcid{0000-0002-4403-3846}
\affiliation{%
	\institution{Jordan University of Science and Technology}
	\country{Jordan}
}

\renewcommand{\shortauthors}{I. Al Ridhawi et al.}

\begin{abstract}
  The rise of fast communication media both at the core and at the edge has resulted in unprecedented numbers of sophisticated and intelligent wireless IoT devices. Tactile Internet has enabled the interaction between humans and machines within their environment to achieve revolutionized solutions both on the move and in real-time. Many applications such as intelligent autonomous self-driving, smart agriculture and industrial solutions, and self-learning multimedia content filtering and sharing have become attainable through cooperative, distributed and decentralized systems, namely, volunteer computing. This article introduces a blockchain-enabled resource sharing and service composition solution through volunteer computing. Device resource, computing, and intelligence capabilities are advertised in the environment to be made discoverable and available for sharing with the aid of blockchain technology. Incentives in the form of on-demand service availability are given to resource and service providers to ensure fair and balanced cooperative resource usage. Blockchains are formed whenever a service request is initiated with the aid of fog and mobile edge computing (MEC) devices to ensure secure communication and service delivery for the participants. Using both volunteer computing techniques and tactile internet architectures, we devise a fast and reliable service provisioning framework that relies on a reinforcement learning technique. Simulation results show that the proposed solution can achieve high reward distribution, increased number of blockchain formations, reduced delays, and balanced resource usage among participants, under the premise of high IoT device availability.
\end{abstract}




\maketitle

\section{Introduction}
The future of intelligent and on-demand time-sensitive service provisioning relies heavily on system distribution and decentralization. At its early stages, smart city applications have relied on centralized solutions such as the Cloud. Most applications (e.g. healthcare, autonomous driving vehicles, etc.) have offloaded their tasks in terms of computing, storage and data analytics to cloud datacenters and storage sites \cite{thar1}. IoT devices simply acted as data collectors with minimal data filtration and analysis at their end. This was mainly due to the IoT devices’ minimal hardware, software and intelligence capabilities. Moreover, cellular communication was restricted by low bandwidth availability, slow data rates, and minimal simultaneous device connections. Most devices relied on short distance communication such as wireless local area networks (WLANs) or wired connections. Such a solution seemed to be promising at first, given the low number of IoT devices. But with the enormous expansion in the number of IoT devices, in addition to the advancements in user device capabilities in terms of hardware, software and communication, traditional cloud solutions were no longer attractive, especially for time-sensitive applications such as autonomous self-driving vehicles \cite{self-driving}, intelligent health monitoring \cite{smartHealthCare}, and emergency response services \cite{emergencyResponse}.

At the early stages of research and development of the Fifth Generation (5G) communication network \cite{aloqaily2020design}, but before its early stage deployment, alternative distributed and decentralized solutions were made available to support time-sensitive applications, namely fog and mobile edge computing (MEC) \cite{fog1}. Fog and MEC were introduced to provide processing and storage solutions located in the vicinity of mobile and edge IoT devices. This will somewhat eliminate latencies and communication delays experienced from the reliance on processing and storage cloud entities. Although such solutions were first experienced with cloudlets \cite{cloudlet}, fog and MEC have seen enhanced performance with today’s smart city and IoT ecosystems. To put things into perspective, fog computing and MEC provide an alternative central access point for not only communication purposes, but also processing, storage, and intelligence. For instance, data that requires immediate attention and processing is handled by the fog, otherwise the job will be offloaded to the cloud. Alternatively, jobs submitted to the cloud can be offloaded to a number of fogs for processing or storage. In both cases, this relieves the pressure off the cloud datacenter and ensures that all jobs are delivered within the requested quality of service (QoS) and quality of experience (QoE) requirements \cite{profitable}.

As technology has progressed at both the communication (e.g. 5G, MANETs, VANETs, etc.) and service (e.g. fog and edge devices) layers, IoT and smart city applications started relying heavily on decentralized and distributed solutions. As such, the concept of fog-to-cloud (F2C) communication shifted more towards fog-to-fog (F2F) communication \cite{profitable}. Both resource sharing and collaboration for task completion became a necessity in order to complete jobs on time and meet the QoS and QoE requirements. Data replication and service availability at different fog sites made it possible for user-specific services to be composed on demand \cite{Ridhawi2018ACM}. Solutions for data and service decomposition were developed to ensure that most of the data and simple services are available at fog sites, thus allowing for services to be composed \cite{Ridhawi2018ACM}. Users requesting services that once were available on the cloud, can now be composed and delivered in a timely manner. More complex services, especially ones that require machine learning techniques, still use resources of cloud datacenters and storage sites. Profit sharing mechanisms were also introduced to motivate cooperation and collaboration among fog and MEC devices that belong to different internet and network service providers (INP/NSP) \cite{profitable}.

Although this provided an opportunity to overcome significant issues at earlier stages of the cloud distribution strategy, various user-specific requests (which arise as a result of new technology availability) still cannot be fulfilled at both the edge and cloud. For instance, multimedia user-specific services that require the rendering of content (e.g. video and audio enhancements, color effects, and language support \cite{workflow}), and which may not be available at the fog or cloud (or at least the added rendering capabilities), can only be supported through end-device cooperation (i.e. resource, hardware and software capability sharing). A significant number of today’s service requests, and most of tomorrow’s requests will require some type of artificial intelligence (AI) integration to achieve enriched service capabilities. As such, reliance on the fog and cloud to fulfill all complex and composite service requests is no longer tolerable. Involvement of resource-rich IoT devices in the service composition and delivery process is thus highly essential \cite{resourceRich}. Decentralization, distribution, collaboration and cooperation, and resource and intelligence sharing at the end-device level is required more than ever to not only achieve the requirements of most user- and device- specific service requests, but also to achieve a balanced workload on all participants in today’s complicated networked ecosystem. Service requesters are no longer acquirers of consumable services, but rather are involved heavily in the service provisioning process. As such the concept of \textit{volunteer computing} in a tactile internet environment is highly needed for tomorrow’s beyond 5G infrastructure, namely, 6G.

Blockchain technology is being employed with several applications and integrated with other technologies such as IoT, health, energy, Fog/Edge, to name few \cite{tseng2020blockchain}\cite{bc1}\cite{bc2}. The vision of 6G is to achieve total connectivity of intelligent things on-ground, in the sea, in the sky, and in space \cite{6g}. To do so, we cannot simply rely on solutions that assume devices will cooperate and collaborate to share their resources and capabilities in the service provisioning process. \textit{Incentives} in any form must be given to participants to ensure fair usage and proper compensation for their involvement \cite{incentive5G}. Such incentives may be in the form of service provider profit sharing or prioritized access to subscribed services. As such, the significant majority of tasks will be conducted by collaborative IoT devices with the aid of fog and MEC computing devices. Centralized entities such as cloud datacenters will act as a backbone to smart city applications and support the decentralization process by offering intelligent processing and storage capabilities for very complex tasks that cannot be delivered at the edge. Securely communicating data and collaborating to form and deliver services can only be achievable with the aid of secure decentralized infrastructures that support device to device communication. A plethora of applications will benefit from such incentivized cooperative solutions. For instance, connected vehicles can share resources (e.g. computing, storage, power, etc.) as part of the cooperation process. In return, service providers get rewarded for such on-demand requests. Multimedia content sharing is another hot topic in today's social networking \cite{fog1}. In densely crowded environments, such as stadiums, spectators can have on-demand access to a game replay with certain user-specific content enhancements from other spectators' devices. Such a collaborative environment will require some sort of an incentive model to ensure cooperation among participants.

This article proposes a cooperative and collaborative solution among edge IoT devices to share their resources, computation, storage and intelligence capabilities. Blockchain is used as a form of decentralization to compose and deliver composite services securely and privately \cite{blockchainICC}. Incentives are provided to participants to ensure that the developed framework is sustainable for all participants, namely, both the service providers and requesters. The contributions of the proposed work are summarized as follows:

\begin{itemize}
	\item A cooperative IoT framework that supports volunteer computing at the end-user device level to share their resources, processing, storage and intelligence capabilities.
	\item An incentive-based mechanism is adopted to the framework to support and offer compensation for participating in resource sharing and service composition processes.
	\item A blockchain technique is adapted to support decentralized service composition and delivery to sustain data privacy and consensus among participants.
\end{itemize}

The rest of the article is organized as follows: Section \ref{relatedWork} explores some of the most recent related work in the literature. Section \ref{problem} formulates the problem and discusses the optimization aspect of the problem. The proposed IoT framework is considered in Section \ref{framework}. Reward distribution and blockchain formation specifics are discussed in Section \ref{reward distribution}. Section \ref{simulations} provides details in regards to the conducted simulations. Finally, we conclude the article in Section \ref{conclusion} with some future work insights.

\section{Related Work}
\label{relatedWork}

The reliance on information and communication technology resources have grown exponentially, and data processing became the most important strategic resource. This is mainly due to the increase in data volume. The solution to big data processing can be achieved through collaboration processing and data sharing, however, unwillingness to collaborate, the fear of resource sharing (i.e. trust issues), and sometimes inability to share due to connectivity problems are few examples of persisting problems \cite{8456052}.

Resource sharing techniques through collaboration (i.e. volunteer computing) are emerging as a mean to reduce repetitive tasks, faster computing, and promote the concept of decentralized open-source computing. However, we still face two main issues in regards to how to properly incentivize participants to volunteer their resources and how to ensure the security and privacy for the participants' data. The work in \cite{xuan2020incentive}\cite{shen2020blockchain} have proposed a mechanism of incentives through advancing the concept of data sharing based on blockchain and smart contracts. The former uses smart contracts to encourage users to share their data to overcome trust fear. Nash game equilibrium analysis was used for incentives. Moreover, a reliable collaboration model for resource owners, miners, and trusted third parties have been proposed in \cite{shen2020blockchain}. The participant signs a smart contract first and then shares data and resources via blockchain. The incentive concept is being developed through revenue distribution among participants.

Volunteer resources are not only limited to regular cloud participants but have been also investigated at the vehicular network level. A privacy-preserving mechanism using blockchain for incentive announcement for communications between connected vehicles, namely, CreditCoin, has been proposed in \cite{8267113}. An aggregation protocol has been used for incentives. Similarly, the authors in \cite{amjid2020vanet} proposed a new paradigm by merging vehicular ad-hoc network (VANET) with volunteer computing to provide efficient utilization of idle computing resources. The authors evaluated their proposed paradigm using job completion, throughput and latency to show better results compared with traditional approaches. Neither incentives nor blockchain were considered in their solution.

When it comes to personal volunteer computing, in \cite{lavoie2019personal}, the authors proposed a distributed computing approach that leverages personal devices (smartphones and laptops) to the personal computation needs from the general public of programmers to perform significant applications or community interest. Such a paradigm, also, encourages developers to maintain and enhance new applications part-time, where no additional hardware is needed, and the process of tools was done over existing devices. While in \cite{mengistu2019volunteer}, the authors have proposed volunteer computing as a service (VCaaS) based edge computing, where volunteer computing resources are employed for edge computing to process data from IoT devices. Security and privacy were considered as well.

To evaluate the effectiveness of such a new model, some researchers have studied the requirements in addition to the strength of current volunteer computing platforms \cite{durrani2014volunteer}. The authors have analysed multiple issues such as the effectiveness of the active participants and how the computation and communication can be performed in addition to the analysis of task distribution and result validation polices. On the other hand, the authors in \cite{nov2010volunteer} addressed the gap between the computational pillars. The authors drew on social psychology and online communities' researches and proposed a three-dimensional model of the factors determining contributions of volunteer computing users (tenure, personal motivations and team affiliation). Also, the authors identified the relations between these factors and the actual contribution level.

Among the many technical challenges emerging from this new technology, the most challenging problem is task scheduling, where the resources are not only heterogeneous but also may go offline at any moment. The authors in \cite{xu2019dynamic} proposed a deadline preference dispatch scheduling (DPDS) algorithm which is based on a dynamic task scheduling algorithm for heterogeneous volunteer computing platforms. In DPDS, the task that has the minimum deadline constraint will complete first by assigning it to a near volunteer node. Also, to maximize the number of computed tasks before the deadline constraint and to fully utilize volunteered resources, the authors proposed an improved dispatch constraint scheduling algorithm (IDCD) where tasks are selected according to their priorities. The authors used a risk prediction model in the IDCD algorithm to ensure efficient application execution by predicting a completion risk of each task. In \cite{parkhomenko2019scheduling}, the authors used a neural network mechanism to predict the job execution time and genetic algorithm, in order to distribute jobs to volunteers with adjusting parameters to make it responsive to any changes. The results showed the benefits of the proposed model even when volunteer computing network dimensions are not high.

As seen from the literature work, several challenges arise as a result of adopting this technology at a large scale in terms of the heterogeneity of the resources, the variety of the capabilities, the distribution of the tasks, the efficiency and utilization of volunteered resources. However, even if all of these issues were managed and solved, retaining a large number of participants' resources, encouraging data sharing, and guaranteeing continuous contributions are only possible if we provide trust and proper incentives. This has not been explored yet, and we believe that this article will address those issues.

\section{Problem Formulation}
\label{problem}
\begin{figure}[h]
	\centering
	\includegraphics[scale=0.37]{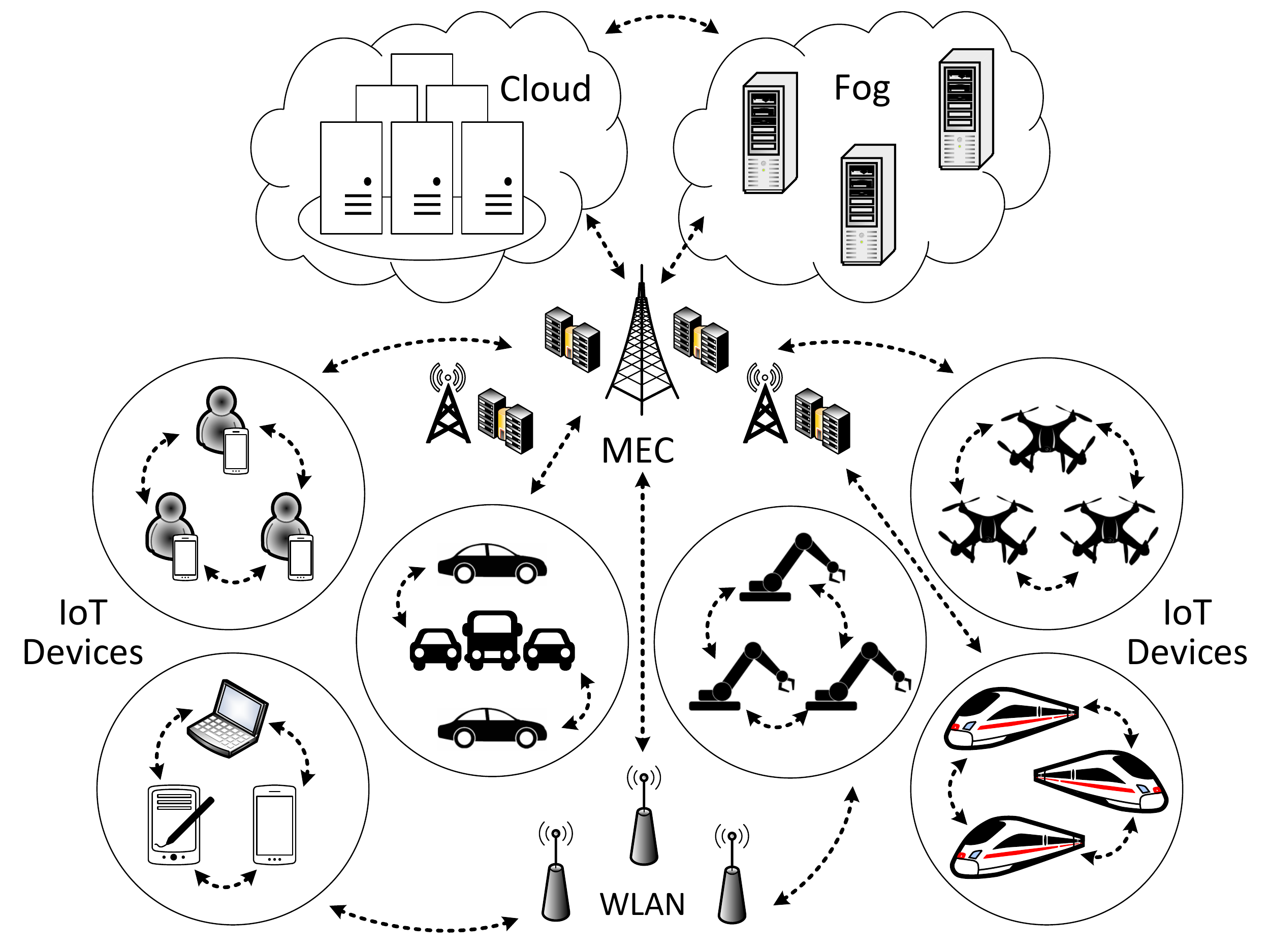}
	\caption{Tactile internet network environment comprising of a plethora of communication technologies, in addition to volunteer computing.}
	\label{fig:tactile}
\end{figure}

We consider a tactile internet network environment, as depicted in Figure \ref{fig:tactile}, comprising of a plethora of access points (APs) and base Stations (BSs) belonging to different ISPs and NSPs of different technologies, such as LTE eNB, Wi-Fi APs, MEC servers, etc. Moreover, the network environment comprises a number of IoT end-devices $UE = \{ue_1,ue_1,...ue_n\}$ that have a set of resources and capabilities defined as $Cap=\{cap_1,cap_2,...,cap_w\}$, and have been requested to cooperate in order to complete a service request $Req_i=(D_i,Q_i,O_i)$. The request is defined through its description properties $D_i$ (defined later), the acceptable levels of specific QoS parameters $Q_i$, and any other requirements $O_i$ such as cost or prioritized preferences. A service is composed of a set of tasks $S_j=\{t_1,t_2,...t_m\}$, in which each task $t$ has a size $\alpha_{t_m}$, dependency on other tasks or sub-tasks $\beta_{t_m}$, and completion deadline $\gamma_{t_m}$. The complexity of a task will be measured by the participant in terms of its computation or storage intensity $\delta_{t_m}$, defined in the form of processor cycles per data block and energy consumption $\zeta_{t_m}$. Such a measurement is not only dependent on the task's characteristics, but also varies in accordance to each node's capabilities (e.g. hardware, software, etc.). 

\subsection{Maximizing Participant Gain}
Each participant $ue_n$ in the service composition and delivery process aims at maximizing its participation gain $G_{t_{n,m}}$ as defined in (\ref{eq1}), by increasing its reward $R_{t_{n,m}}$ as defined in (\ref{eq2}), reducing its workload $W_{t_{n,m}}$ as defined in (\ref{eq3}), and eliminating/decreasing its penalties $P_{t_{n,m}}$ as a result of performing task $t_m$ as defined in (\ref{eq4}). Tasks with a large size $\alpha_{t_m}$, more dependencies $\beta_{t_m}$, and stricter completion time $\gamma_{t_m}$ will lead to higher rewards $R_{t_{n,m}}$. Furthermore, achieving the requested QoS levels for a task $q_{t_m}$ within the set time limits will lead to higher reward values. On the contrary, not achieving the task (or achieving the task but not meeting the requested QoS and time limits) will lead to more penalties $P_{t_{n,m}}$, hence, less rewards. Penalties are incorporated within the gain function to ensure that participants are performing the requested tasks on time and in accordance to the set service requirements. Nodes that simply join the composition process without adhering to the set rules (i.e. nodes participating in compositions beyond their resource capabilities for large reward returns) will receive penalties for not adhering to the set service requirements. This will ensure that fair participation among cooperating nodes is achieved.
Moreover, the workload $W_{t_{n,m}}$ is dependent on the participants capabilities, such that an end-device that is described as energy-efficient with high and complex processing capabilities will complete a task with less workload (e.g. time, power usage, etc.). The requested/expected levels for reward, workload and penalty specifications, namely, $\chi_{t_m}=(q_{t_m},\gamma_{t_m},\delta_{t_m},\zeta_{t_m})$ is compared to the actual levels achieved at time $t$, namely, $\grave{\chi_{t_m}}(t)=(\grave{q_{t_m}}(t),\grave{\gamma_{t_m}}(t),\grave{\delta_{t_m}}(t),\grave{\zeta_{t_m}}(t))$. Having a solution where $\grave{\chi_{t_m}} \geq \chi_{t_m}$ results in higher rewards and less workload, which in essence leads to higher gains $G_{t_{n,m}}(\chi_{t_{m}},\grave{\chi}_{t_{m}}(t))$.

\begin{equation}\label{eq1}
G_{t_{n,m}}(\chi_{t_{m}},\grave{\chi}_{t_{m}}(t)) = R_{t_{n,m}}(\chi_{t_{m}},\grave{\chi}_{t_{m}}(t)) - W_{t_{n,m}}(\chi_{t_{m}},\grave{\chi}_{t_{m}}(t)) - P_{t_{n,m}}(\chi_{t_{m}},\grave{\chi}_{t_{m}}(t))
\end{equation}

\begin{equation}\label{eq2}
R_{t_{n,m}}(\chi_{t_{m}},\grave{\chi}_{t_{m}}(t)) = \sum\limits_{m=t_0,\underline{q}_{t_{m}}\leq{\grave{q}_{t_{n,m}}(t)}\leq\overline{q}_{t_{m}}}^M \tau_q \grave{q}_{t_{n,m}}(t) + \sum\limits_{m=t_0,\grave{\gamma}_{t_{n,m}}(t)\leq \gamma_{t_{m}}}^M \tau_\gamma max\left ( 0,(\gamma_{t_{m}} - \grave{\gamma}_{t_{n,m}}(t)) \right )
\end{equation}

\begin{equation}\label{eq3}
W_{t_{n,m}}(\chi_{t_{m}},\grave{\chi}_{t_{m}}(t)) = \sum\limits_{m=t_0}^M  \grave{\delta}_{t_{n,m}}(t) + \sum\limits_{m=t_0}^M \grave{\zeta}_{t_{n,m}}(t)
\end{equation}

\begin{equation}\label{eq4}
P_{t_{n,m}}(\chi_{t_{m}},\grave{\chi}_{t_{m}}(t)) = \sum\limits_{m=t_0,\underline{q}_{t_{m}}\geq{\grave{q}_{t_{n,m}}(t)}\geq\overline{q}_{t_{m}}}^M \sigma_q \grave{q}_{t_{n,m}}(t) + \sum\limits_{m=t_0,\grave{\gamma}_{t_{n,m}}(t)\geq \gamma_{t_{m}}}^M \sigma_\gamma max\left (0,(\grave{\gamma}_{t_{n,m}}(t)-\gamma_{t_{m}}) \right )
\end{equation}

$\tau_q$ is the reward given for a participant that successfully completes the task within the quality limits set, namely, $\underline{q}_{t_{m}}\leq{\grave{q}_{t_{n,m}}(t)}\leq\overline{q}_{t_{m}}$. The floor and ceiling QoS values are prone to change frequently according to network performance. If the network resources are limited, then the offered QoS value range for a particular service would be reduced to accommodate for the available resources (i.e. participant resources). Details in regards to dynamic configurations of network parameters is out of the scope of this article and has been discussed in \cite{profitable}.
$\tau_\gamma$ is the reward per time unit given for participants that successfully complete the assigned task at the deadline time $\gamma_{t_{m}}$, such that higher rewards are given for less time units $\grave{\gamma}_{t_{n,m}}(t)$ needed to complete the task. $\grave{\delta}_{t_{n,m}}(t)$ is the processor/storage workload incurred on the participant for performing the given task which resulted in the consumption of processor/storage resources at time $t$. Similarly, $\grave{\zeta}_{t_{n,m}}(t)$ is the workload incurred on the participant for the power consumed to perform the given task. The evaluation of the proposed system’s energy consumption is measured in terms of the nodes workload to complete a task. The computation intensity (i.e. CPU cycles per bit) to perform a service task is considered when analyzing the energy consumption for a candidate participant. We adopted the technique introduced in \cite{energyConsumption} to measure the power usage per CPU cycle. $\sigma_q$ is the penalty incurred on the participant for performing the given task which resulted in QoS values below the requested levels. Similarly, $\sigma_\gamma$ is the penalty incurred on the participant for not meeting the task deadline $\gamma_{t_{m}}$.

As such, the objective that must be considered while distributing tasks among participants is to maximize the gain achieved among all participants while adhering to obligations arising from the requested service as described in (\ref{eq5}). 

\begin{equation}\label{eq5}
\begin{array}{l}
\boldsymbol{P1}:\left ( maximize \sum\limits_{n=1}^N \sum\limits_{m=t_0}^M G_{t_{n,m}}(\chi_{t_{m}},\grave{\chi}_{t_{m}}(t)) \right )
\\ s.t. \quad \boldsymbol{C1}: \quad max \sum\limits_{n=1}^{N_{t_{m}}\subset N} C_n
\end{array}
\end{equation}

The optimization problem is solved for collaboratively by all nodes whom are willing to participate and collaborate to deliver simple and composite services. Details in regards to the solution is looked at in Section \ref{reward distribution}. The selection process is reliant on the rank given by other participating nodes towards the participant's behaviour, which is dependent on previous successful task completions and the node's cooperation willingness characteristics. Hence, as seen in the constraint of (\ref{eq5}), the set of participants which attain the maximum node cooperation and willingness score ($C_n$), defined in Section \ref{framework}, are selected for the composition and delivery process.

\subsection{Blockchain Formation}
The other issue that needs to be considered is the blockchain formation problem. Not only participants' capabilities and their scores are considered, but also whether the result of performing a task by the participant is consistent with the input of the next block in the blockchain, hence, blockchain formation. The selection process must ensure that similarity measures between sequential blocks are considered in the formation process. The goal is to formulate a composition path (i.e. blockchain) that increases the similarity score between one block and another, in addition to the overall transaction for a service request. We consider the following optimization problem as defined in (\ref{eq6}). 

\begin{equation}\label{eq6}
\begin{array}{l}
\boldsymbol{P2}:\left ( max \sum\limits_{n=1}^{N_{t_{m}}\subset N} \sum\limits_{m=t_0}^M 
Comp_{char}(ue_{n,t_{m}},ue_{n+1,t_{m}+1})
%Sim_{t_{m}}(B_n,B_{n+1}) 
\right )
\\ s.t. \quad \boldsymbol{C1}: \quad max \left( RW_{B_{comp}} \right)
\end{array}
\end{equation}

\noindent where $RW_{B_{comp}}$ is the reward value for a service composition transaction using blockchain, attained through a reinforcement learning algorithm. The reward value is determined as a result of previous records of blockchain formations' experiences resulting from the selection of different blockchain patterns.

\section{Cooperative IoT Framework}
\label{framework}
Participants in the service sharing and composition process have two strategies, namely, \emph{participate} or \emph{not-participate}. The strategy is dependent on a number of factors: $i)$ the device's \emph{capability} (i.e. hardware, software, etc.), $ii)$ the user's \emph{cooperation rationality} given certain participation constraints, $iii)$ the device's \emph{cooperation awareness} in terms of continuous learning ability through analysis and strategy readjustment. 

\subsection{Participant Capabilities}
\label{capabilities}
Upon joining a network environment as a participant in the service provisioning process, end-devices ($ue_n$) communicate their capabilities $Cap_{ue_n}$ set (see Example 1, presented in XML format for reader-friendly purposes) to the nearest MEC device either directly through point-to-point communication or through other devices such as WiFi APs, device-to-device communication for Ad Hoc networks, etc. The capability list and directory is consistently updated and shared among all fog and cloud entities. Whenever a service request $Req_i$ is communicated to the participants with a defined set of description properties $D_i$, described through an ontological structure \cite{fuzzy1}, participants compare the service request properties against their capabilities. Nodes can determine whether the needed resources/capabilities are available using syntactic and semantic similarity comparison against the request \cite{ontology3}. Details are out of the scope of this article and have been covered in an earlier work \cite{ontology3}. 

\begin{figure}[h]
	\centering
	\fbox{\includegraphics[scale=0.58]{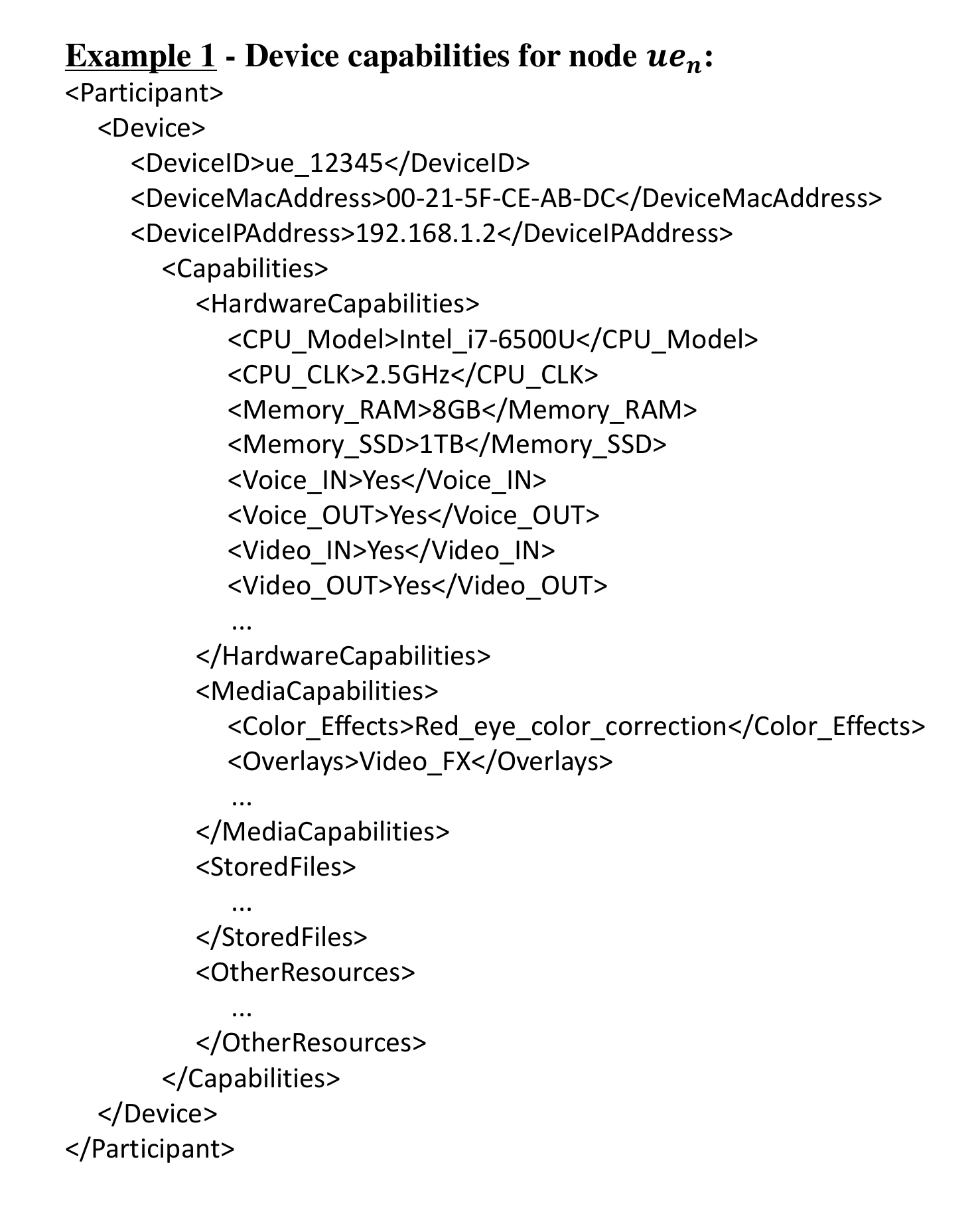}}
	%\caption{Tactile internet network environment comprising of a plethora of communication technologies, in addition to volunteer computing.}
	\label{fig:device_capabilities}
\end{figure}

\subsection{Participant Cooperation Rationality}
As described earlier, not only the intent for participants to maximize the gain from the cooperation is necessary, but also to ensure that highly capable and cooperative nodes are joining the cooperation process to ensure diversified resource and service availability. A participant's rationality towards cooperation is dependent on a different factors, namely, $i)$ the type and characteristics of the service request, $ii)$ it's current behaviour towards cooperative entities based on the current network status and previous experiences, and $iii)$ the participant's cooperative status.

\subsubsection{Participant Task Characteristic Preferences}
\hfill\\
Other than the device's capabilities towards achieving a service task, participant's may decide to join or not join the cooperation process due to the type and characteristics of the service. Participants may have certain preferences in terms of what service tasks it may want to perform. For instance, a participant may decide to not join non-educational service requests or that of another characteristics. As such, upon joining a network environment, participants advertise their service characteristics participation preferences and priorities to ensure that they are excluded from non-preferred services. Example 2 provides an example of such advertisement. 

\begin{figure}[h]
	\centering
	\fbox{\includegraphics[scale=0.52]{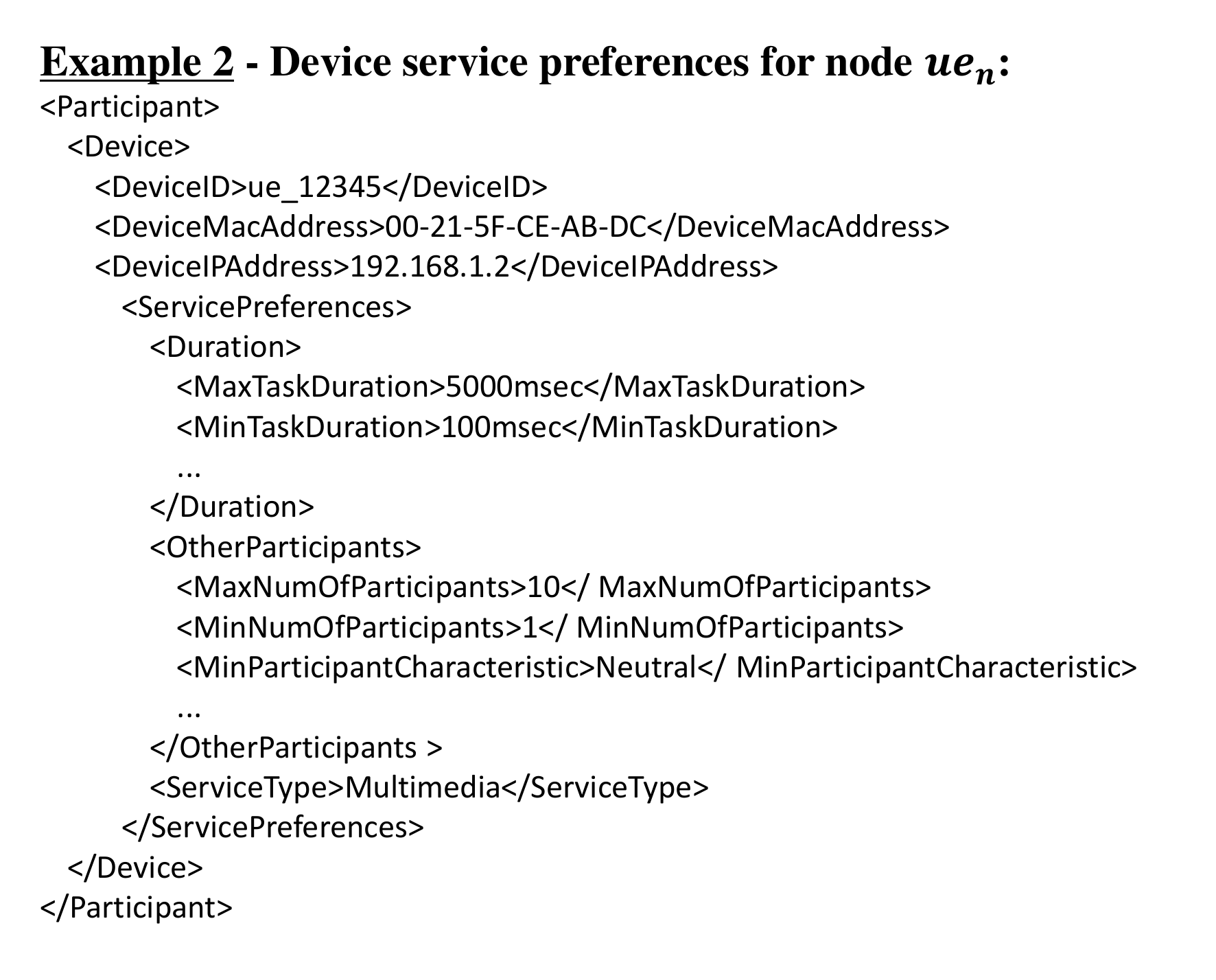}}
	%\caption{Tactile internet network environment comprising of a plethora of communication technologies, in addition to volunteer computing.}
	\label{fig:device_capabilities}
\end{figure}

Task characteristics and preferences are stored in a hierarchical ontology structure, modeled as objects, concepts and relationships. As such, variations in service semantics and functionality syntax are compared to determine similarity and differences among objects. Thus, task characteristics are compared against participant preferences to determine matching and non-matching features. More matching features in comparison to non-matching features determines that a service preference match is likely found. On the contrary, whenever a task is described to be not in the preference category of the participant, then less matching features are found in comparison to non-matching features in the task and participant preferences characteristics. This concept is modeled and defined in (\ref{eq7}).

\begin{equation}\label{eq7}
Comp_{char}(ue_{n},t_{m}) =  \frac{\left | char_{ue_{n}}^{pref} \cap char_{t_m} \right |}{w_{match} \left | char_{ue_{n}}^{pref} \cap char_{t_m} \right | + w_{match} \left | char_{ue_{n}}^{pref} \cup char_{t_m} \right |}  
\end{equation}

\noindent where $\left | char_{ue_{n}}^{pref} \cap char_{t_m} \right |$ is the number of matching features when comparing the participant's task preferences against the requested service task characteristics. Similarly, $\left | char_{ue_{n}}^{pref} \cup char_{t_m} \right |$ represents the non-matching features in the comparison.

\subsubsection{Participant Behaviour}
\begin{figure}[h]
	\centering
	\includegraphics[scale=0.52]{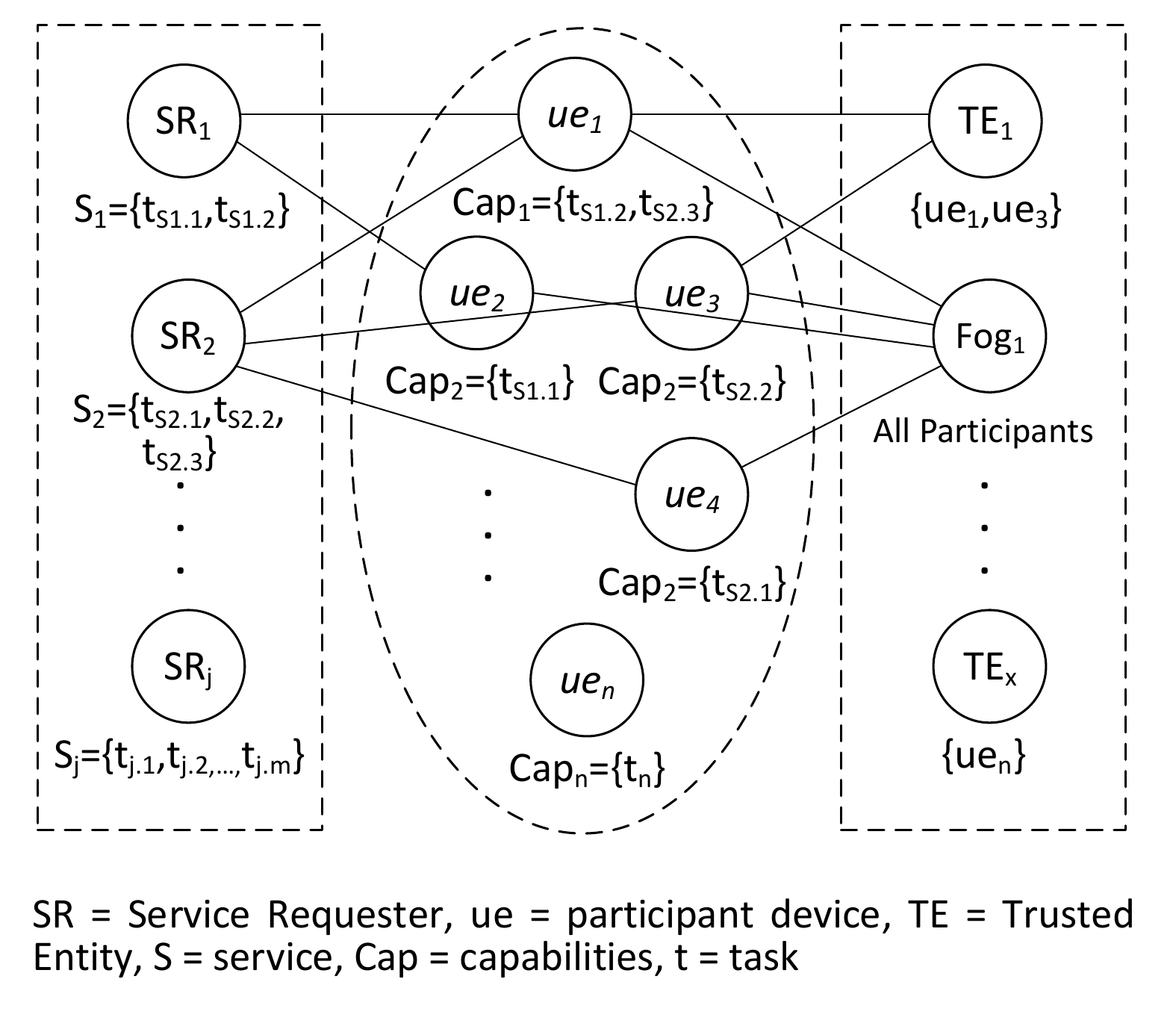}
	\caption{Requests in the form of service tasks are fulfilled using the capabilities of participant devices. Each participant is ranked by the service requesters and fog/trusted entities that have knowledge about the behaviour of participants, based on the current service composition process and/or any previous compositions.}
	\label{fig:ranking}
\end{figure}

\hfill\\
The participant's behaviour towards other participants in the cooperative network is very crucial in determining which set of devices will provide the optimal solution collaboratively. The behaviour of participants is dependent on different criteria, but most importantly the nature of the user. For instance, a participant that has had negative feedback as a result of cooperation with other participants categorized under a certain characteristics category may pose strict conditions for cooperation for future service requests. Therefore, a ranking strategy is adopted to classify the participants' behaviour. All participants involved in the cooperation process will rank each other at the end of the service composition task, in addition to the serving MEC/Fog device and trusted entities, as depicted in Figure \ref{fig:ranking}. This will ensure that a fair score is given to all participants and allows only those with an acceptable level of behaviour join the cooperation process and share the distributed rewards. The figure outlines an example of where service requesters $SR_j$, fog service providers, and trusted entities $TE_x$ rank participants (whom volunteer their service capabilities $Cap_n$) in accordance to their behaviour, based on current and previous service composition processes. Ranking by requesters and trusted entities is only performed by those with direct cooperation towards participants, whereas the fog ranks all participants.

We categorize participants' behaviours using the following ranks/categories:
\begin{itemize}
	\item \textit{Highly Non-Cooperative} - Participants of this category are described as devices that pose security threats to the network, regardless of whether the participant has been asked to join a composition task or not. Such participants need to be identified, reported, isolated and removed from the network to avoid any future threats to the network structure.
	\item \textit{Non-Cooperative} - Such participants pose a security and/or privacy threat to the network only when requested to join a cooperative composition task. Hence, such devices need to be identified, reported and isolated, but not removed from the network environment. Nodes of this type may request services but cannot participate in the service provisioning process.
	\item \textit{Neutral} - Nodes described as neutral pose no security issues to the network and may participate in the cooperation process. The results of whether the participant will sufficiently attain the task requirements are unknown, and thus should only be requested to join the composition process when no other participant of a higher rank is available. Devices of this type are usually casually interested in attaining rewards from cooperation tasks.
	\item \textit{Partially Cooperative} - Participants of this category have a fluctuating level of cooperative task participation which is dependent on the type and characteristics of the service request, in addition to the network conditions. The objective of such participants is not only the amount of participation reward, but also the context of the service.
	\item \textit{Cooperative} - Such participants are considered cooperative at all times and the main objective of such nodes is to increase the reward value. The participant may sometimes decline the service task request whenever the penalties and workload outweighs the rewards. Therefore, it is very important that rewards given for such participants are attractive enough to ensure such nodes join the cooperation process.
	\item \textit{Highly Cooperative} - This rank categorizes participants as highly cooperative in the sense that in almost all circumstances, devices of this type will join the cooperation process even if sometimes the task may lead to a loss in gain. Participants will join all cooperation tasks unless a security or privacy threat is posed by other participants.
\end{itemize}

We adopt a weighted behaviour fuzzification function developed in \cite{fuzzy1} to give a score for each category. This allows for the dynamic adjustment for the participants' categories and provides accurate and reliable participant scores. The fuzzified participant's cooperative behaviour is defined in (\ref{eq8}).

\begin{equation}\label{eq8}
C_n=z_{category}\left (\sum\limits_{n=1}^{N_{t_{m}}\subset N}score_{t_{m}}(ue_{n},ue_{n+1}) \right )  
\end{equation}

\noindent where $score_{t_{m}}(ue_{n},ue_{n+1})$ is the score in terms of cooperative characteristics that each participant end-device, fog and trusted entity provides to all other participants. $z_{category}()$ is the fuzzification function used to determine the fuzzified participant's cooperative behaviour $C_n$.

\subsubsection{Participation Status}
\hfill\\
Although incentives are provided to share resources and get involved in the composition process. Serving fog and MEC entities have authority to ban certain nodes not only according to their cooperation category, but also in accordance to their current and previous participation status. For instance, nodes that show greedy behaviour, where they only participate whenever highly valuable rewards are given may be banned from participation in future cooperation sessions to ensure fair distribution of rewards to other novice participants. Moreover, a node itself may decide to whether participate in cooperation/composition tasks or not. As such the participation status changes in accordance to the participant's desire, in addition to the context. We also note here that serving fog entities may ban nodes from participating in cooperation tasks if participants reject cooperation requests excessively.

\subsection{Participant Cooperation Awareness}
We assume that most participants have machine learning (ML) capabilities and at the same time can achieve the task of federated learning collaboratively. Each participant adapts its own ML algorithm and relies on the Stochastic Gradient Decent (SGD) method to perform stochastic approximation of gradient descent optimization and replaces the data set gradient with an estimated one through random sub-set data selection \cite{fl1}. In terms of services that require distributed and collaborated federated learning tasks, tasks' outputs and trained models are either authenticated using the blockchain consensus algorithm or directed to the serving fog node for authentication and aggregation. Such a decision on whether to use blockchain consensus or the fog is dependent on the type of service request. Services classified as '\textit{sensitive}' are directed to the fog, otherwise the blockchain consensus method is used. 

Furthermore, with regards to whether to participate or not-participate in the service sharing and composition process, the training on data is performed locally on the end-user devices. Participants use previous gain achievements and other criteria in regards to cooperation of other participants to determine whether future cooperative sessions are ideal in regards to the gains achieved. Local data is also trained to avoid participating with other devices or share data that may be classified as hazardous leading to intrusion attacks \cite{intrusion}.

\section{Reward Distribution}
\label{reward distribution}
Participants are selected in accordance to their advertised capabilities and the need for task distribution among a number of end-devices. As discussed earlier in Section \ref{capabilities}, participants advertise their capabilities to the nearest fog/MEC device. From there, the information is shared among neighbouring fog and MEC nodes. We classify the participant search process into two categories, namely, \textit{simple} and \textit{complex}. The former considers service compositions that rely entirely on the advertised data by participants to fog/MEC nodes, where the participants are selected and a blockchain is formed to record the composition process, which is later used for the reinforcement learning process (discussed later). The latter, requires the aid of miners (i.e. trusted entities) to search for capabilities not registered on the framework and requires coordination among participants and miners to complete the composition process on the blockchain. Moreover, the selection process among the two methods is also dependent on the time-sensitivity and QoS restrictions.

\subsection{Simple Search Process}
\label{simpleSearch}
\begin{figure}[h]
	\centering
	\includegraphics[scale=0.42]{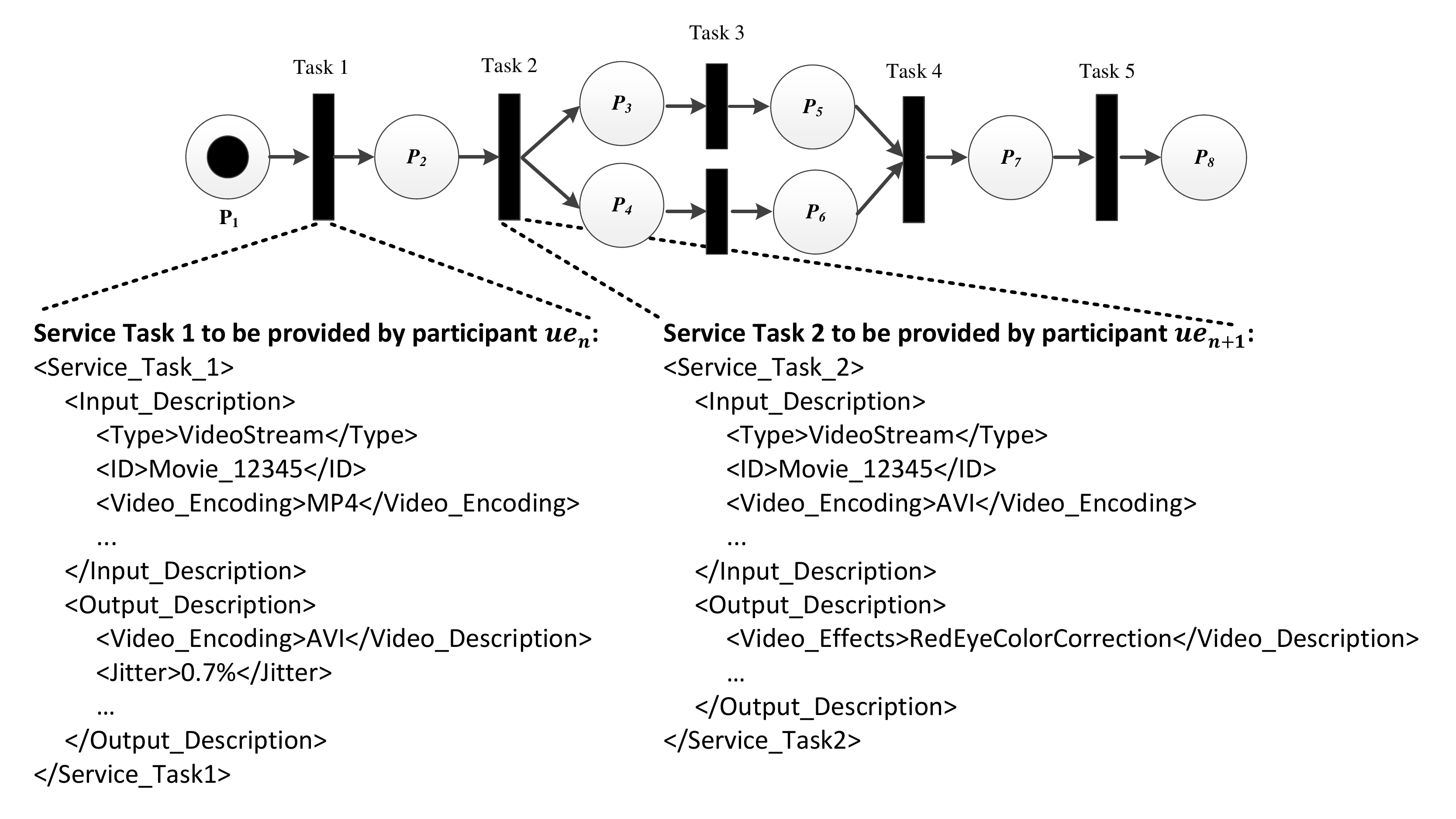}
	\caption{A workflow plan constructed by the serving Fog/MEC device outlining the set of tasks needed (described as a transition) and the participants selected to perform the tasks (described as places) in accordance to their cooperative behaviour score as defined in (\ref{eq8}).}
	\label{fig:workflow1}
\end{figure}

Upon receipt of the service request from a requester for a simple service (i.e. one with restricted time and QoS constraints or with availability of matching capabilities), a workflow plan is constructed to determine the tasks needed to be composed to deliver the composite service request. In addition to the tasks, the workflow plan identifies the best matching candidates in accordance to their cooperative characteristics scores as defined in (\ref{eq8}). Figure \ref{fig:workflow1} illustrates an example of a workflow plan constructed by the serving fog device. In our work, service tasks performed by participants in the composition process are modelled using Workflow-nets which are an extension to Petri-nets \cite{workflow}. Workflows guarantee the correctness of the cooperation and reachability problem \cite{workflow}. A Petri-net is a directed graph in which nodes are either transitions or places, where a place $P$ is connected to one or more transitions represented as \textit{Tasks}. Transitions perform service tasks and are represented as tokens residing in places. A transition is said to be enabled only when there are no empty places (i.e. places with no tokens) connected to it as input. When a transition executes a task, tokens are removed from each of the transition’s input places and tokens are created in each of its output places. Moreover, workflows ensure that there is one place with no incoming transition and one place with no outgoing transition. More details in regards to workflows are highlighted in \cite{workflow}.

Resource and capability acquisition requests are sent along with detailed information in regards to the task description in terms of its size, complexity, dependency, etc... to the selected candidates to identify their willingness to participate in the cooperation process. Each node calculates its anticipated gain (if any) according to (\ref{eq1}) and forwards the information to the serving fog node. In order for the participant to be selected and for the blockchain to be formed, the fog node will select a set of candidates that achieve maximized gains according to (\ref{eq5}). A blockchain is then formed from the participants to ensure that the composition process is guaranteed and recorded. The blockchain formation must consider (\ref{eq6}), where the similarity score between two blocks (i.e. difference in semantic distance) is measured to ensure difference is minimized (i.e. similarity is maximized). A smaller difference in semantic distance represents a beneficial move towards meeting the service requirements, indicating that the service request described through an ontology is nearly matching the outputs provided by the participants. The blockchain formation process adheres to positive results accumulated from previous similar blockchain formation trials. In essence, a reinforcement learning process is adopted \cite{profitable} to determine the reward value ($RW_{B_{comp}}$) defined in (\ref{eq9}) that may be attained from similar blockchain formations and to speed up the formation process.

\begin{equation}\label{eq9}
RW_{B_{comp}}=\sum\limits_{n=1}^{N_{t_{m}}\subset N} \left( P \left (\bar{Comp_{char}}(ue_{n},ue_{n+1}) \right ) \times \tilde{Comp_{char}}(ue_{n},ue_{n+1}) \right)
\end{equation}

\noindent where $P \left (\bar{Comp_{char}}(ue_{n},ue_{n+1}) \right )$ is the probability of achieving the highest similarity for a selected block, $\tilde{Comp_{char}}(ue_{n},ue_{n+1})$ is the expected highest similarity for a selected block. A matrix is formed for all the different blockchains that may be formed and the similarity achieved by each selected block in the blockchain pattern. The value function for selecting a block from a set of alternative blocks in a blockchain is therefore:

\begin{equation}\label{eq10}
V_{ue_{n}}(t)=V_{ue_{n}}(t-1)+\rho \left ( V_{ue_{n}} - V_{ue_{n}}(t-1)  \right )
\end{equation}

\noindent where $V_{ue_{n}}(t-1)$ is the previous value function at time $t-1$, $\rho$ is the learning rate, and $V_{ue_{n}}=RW_{B_{comp}}$. As such, the blockchain which results in the highest value function is selected to ensure that constraint $C1$ is met as defined in (\ref{eq5}).

Upon completion of the block selection process, all selected devices are informed and a blockchain is formed to complete the composition process and record all transactions on the blockchain. The requested service is then delivered to the requester and rewards are distributed to all participants. All participants involved in the blockchain formation process then rank each other, in addition to the serving fog node. The fuzzified participant cooperative behaviour score, defined in (\ref{eq8}) is then updated. 
%Algorithm 1 summarizes the process of blockchain formation in accordance to the \textit{simple} search method.

\subsection{Complex Search Process}
\begin{figure}[h]
	\centering
	\includegraphics[scale=0.8]{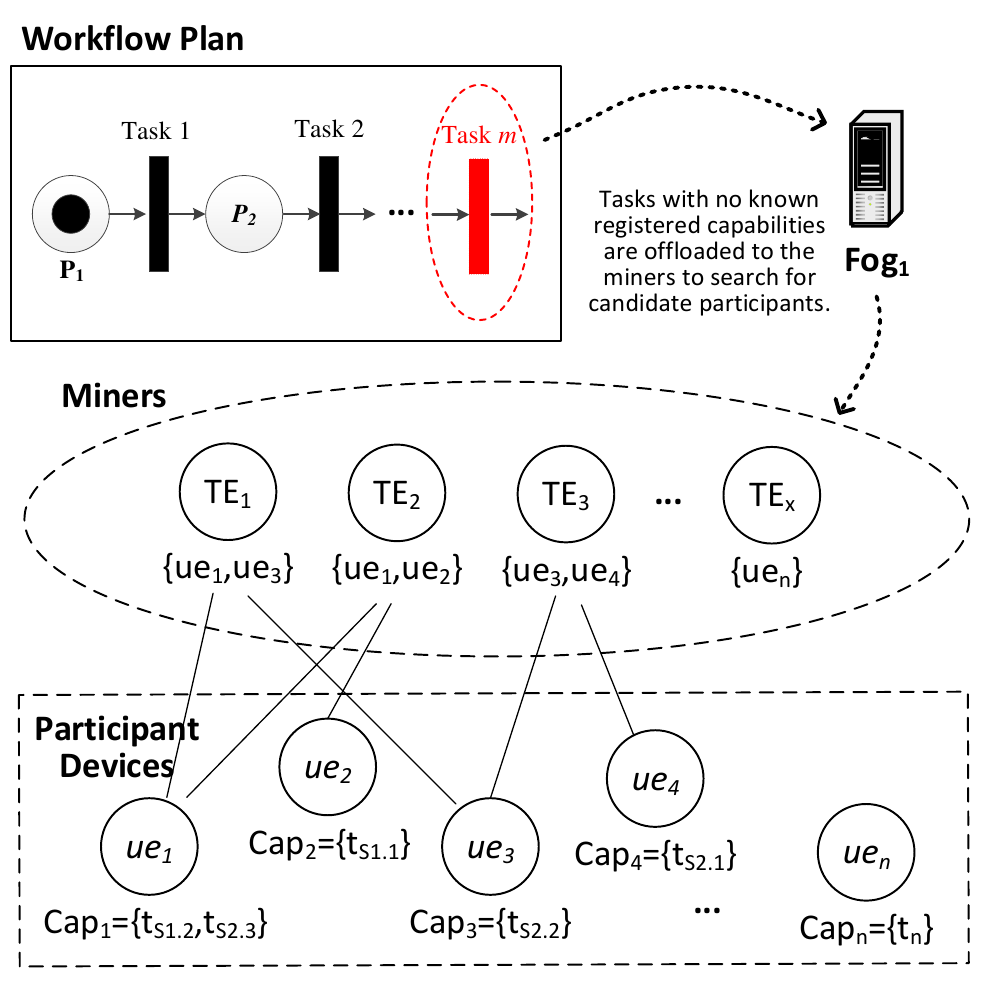}
	\caption{Miners assist in the blockchain formation process by searching for end-devices with capabilities needed to perform service tasks in cases of no registered capabilities or strict QoS requirements.}
	\label{fig:complexSearch}
\end{figure}

For service requests which require capabilities that are not registered among fog/MEC nodes, the search process is considered \textit{complex} and requires the aid of miners to search for capabilities on the framework and coordinate with participants to complete the blockchain formation process. Such a scenario can also be applied to cases with stringent QoS demands but relaxed time-sensitivity to ensure maximized QoS adherence. Upon formation of the workflow plan to determine the needed tasks and capabilities, the process for participant selection with registered capabilities is identical to that of a simple search process (described in Section \ref{simpleSearch}). Participants with matching capabilities are selected as candidate nodes to be part of the blockchain. On the contrary, tasks with no matching registered/advertised capabilities will follow the complex search process. Miners (i.e. trusted entities) are notified of the capabilities needed to construct the block, which in essence will be rewarded for their mining tasks. Miners must also ensure that participants' cooperative characteristics adhere to the constraints defined in (\ref{eq5}) and (\ref{eq6}). Figure \ref{fig:complexSearch} depicts an overview of the complex search process.

Miners are defined as fixed or mobile network and mobile devices that have the capability of communicating directly with other end-devices through different communication methods (e.g. Ad Hoc). End-devices can gain the role of miners once the participant is labeled as trusted. Such a label is given by fogs once the participant's cooperative characteristics score exceeds a predefined fog threshold, namely, $C_n \geq \vartheta$. The threshold $\vartheta$, is a dynamic value that changes in accordance to the network condition. For instance, a network with few participants will have a relaxed threshold value to ensure that the service composition and delivery process is achieved. On the contrary, a highly dense network may have more stringent threshold values to ensure accurate service quality adherence. We assume that the dynamic configuration process follows that of a tabu-search assisted variable configuration optimization mechanism introduced in \cite{profitable}.
As such, end-devices having both roles, namely, participants and miners are capable of increasing there reward significantly. Miners collect rewards for participating in the search process. Tasks which require the aid of miners will have the reward $R_{t_{n,m}}(\chi_{t_{m}},\grave{\chi}_{t_{m}}(t))$ shared among both the miners and selected participant end-devices. The portion of the share is dependent not only on the complexity needed to find the participant, but also finding other participants in the composition process that will provide accurate and stable blockchain formation which adheres to the overall QoS requirements. Therefore, the reward value for miners is determined as a portion $\varphi$ of the reward value determined by the fog as described in (\ref{eq11}). 

\begin{equation}\label{eq11}
R_{t_{TE,m}}=\varphi \left( R_{t_{n,m}}(\chi_{t_{m}},\grave{\chi}_{t_{m}}(t)) \right)
\end{equation}

\begin{figure}[h]
	\centering
	\includegraphics[scale=0.25]{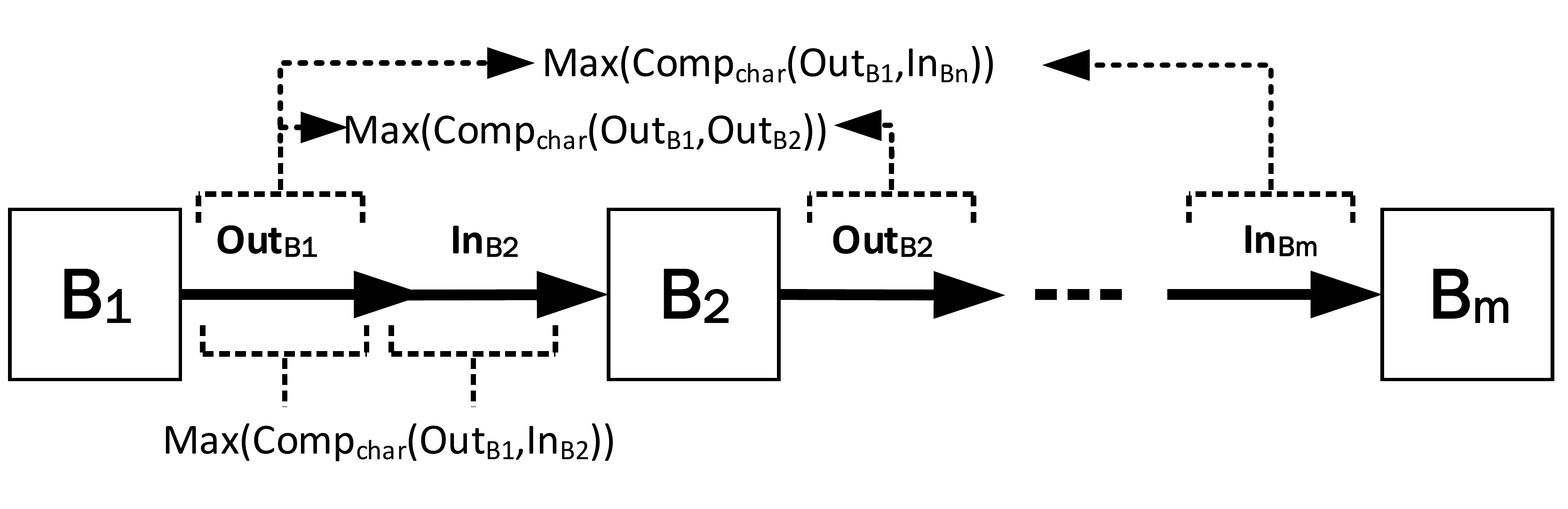}
	\caption{Selecting a set of blocks with the aid of miners to form a complete blockchain.}
	\label{fig:blockchainFormation}
\end{figure}

The blockchain formation process follows the goal of forming a composition path that reduces the semantic distance (i.e. increases similarity) between the current output of the block and that of both the input of the next block and the service request. In essence, a blockchain is formed such that the result of the blockchain (i.e. composition process) increases the semantic similarity with the service request. Figure \ref{fig:blockchainFormation} visualizes the blockchain formation process. From the figure, we see that the output of the first block (i.e. service task performed by participant $ue_1$) and the input of the next candidate block is compared to ensure that the one with the highest $C_n$ value is selected. At the same time, the semantic similarity between the output of the current block and the service request requirement $Req_i$ is compared against that of the output of the next candidate block and the requirements $Req_i$. Such a technique will guarantee that the strict QoS conditions of the requester are delivered.

Additionally, before constructing the blockchain, miners report their candidate participants to the fog, in which the former ensures that the participants in the blockchain achieve maximized gains in accordance to (5). Algorithm 1 summarizes the process of forming a blockchain using the complex search method.

\begin{algorithm}
	\caption{Blockchain formation using the complex search procedure} 
	\begin{algorithmic}[1]
		\State \textbf{Input:} Service request $Req_i$ is sent to the serving fog.
		\\ \textbf{If} {($Req_i$ has stringent QoS $\lor$ $Cap$ missing requested capability $cap_w$)}
		\State Construct workflow plan;
		\\ \textbf{For} {(each task $t_m$ with no $cap_w$)}
		\State Request all available miners $TE$ to search for cap $t_m;$
		\\ \textbf{For}{(each $TE_x$)}
		\State Calculate $Sim_1(TE_x)=Max(Comp_{char}(Out_{B_{n}},In_{B_{n+1}}))$;
		\State Calculate $Sim_2(TE_x)=Max(Comp_{char}(Out_{B_{n}},Out_{B_{n+1}}))$;
		\State Calculate $Sim_3(TE_x)=Max(Comp_{char}(Out_{B_{n}},In_{B_{m}}))$;
		\State Send results $Sim_1(TE_x)$, $Sim_2(TE_x)$, $Sim_3(TE_x)$ to fog;
		\\ \textbf{EndFor}
		\State Determine $ max \sum\limits_{n=1}^{N_{t_{m}}\subset N} \sum\limits_{m=t_0}^M Comp_{char}(ue_{n,t_{m}},ue_{n+1,t_{m}+1})$;
		\State Construct blockchain;
		\State Distribute rewards to all participants according to $R_{t_{TE,m}}$ and $ R_{t_{n,m}}(\chi_{t_{m}},\grave{\chi}_{t_{m}}(t))$;
		\\ \textbf{EndFor}
\\ 		\textbf{EndIf}
	\end{algorithmic} 
\end{algorithm}

\section{Performance Evaluation}
\label{simulations}
Simulations were conducted using OMNET++ \cite{omnet} and OverSim \cite{oversim} as an overlay model for service-specific overlays to mimic blockchains. Private blockchains are deployed with the SHA-256 hash algorithm being used to ensure consistent and secure communication between participants. In fact, multiple private blockchains are created, one for each composition. The size of the simulated network ares was $1000 \times 1000$ meters, with 10 MEC devices and up to 500 end-devices uniformly distributed in the network. The number of trusted entities (i.e. miners) was set to 10\% of the number of participants. All end-devices, including miners act as both service requesters and providers. MEC devices act as 802.11g APs with a bandwidth of 54 Mbps, with both computing and storage capabilities. All end-devices are mobile with a speed of 1-2 meters per second. Service task and capability descriptions are specified in OWL/RDF format \cite{owl}. Capability characteristics similarity evaluations were conducted with the aid of OntoCAT \cite{ontocat}. The fuzzification and reasoning processes were implemented using the jFuzzyLogic fuzzy engine \cite{jFuzzy}. Table \ref{simulationTable} provides a summary of the settings and configurations adapted in the simulator.

\begin{table}[ht]
	\caption{Simulator Settings}\label{simulationTable}
	\centering
	\begin{tabular}{|l|l|} 
		\hline
		\textbf{Simulation Parameters}   & \textbf{Numerical Values}                             \\ 
		\hline
		\textbf{Communication Protocol}  & IEEE 802.11g (for communication between UEs and APs)  \\ 
		\hline
		\textbf{Bandwidth}               & 54 Mbps                                               \\ 
		\hline
		\textbf{Number of APs}           & 10                                                    \\ 
		\hline
		\textbf{Number of UEs}           & 500                                                   \\ 
		\hline
		\textbf{Number of Miners}        & 10\% of UEs                                           \\ 
		\hline
		\textbf{UE Mobility Speed}       & 1-2 m/s                                               \\ 
		\hline
		\textbf{Mobility model}          & Random Waypoint                                       \\ 
		\hline
		\textbf{Blockchain Hash Algorithm}          & SHA-256                                               \\ 
		\hline
		\textbf{Transmission/Idle Power} & 0.1 W/0.01 W (for APs) and 0.02 W/0.001 W (for UEs)   \\
		\hline
	\end{tabular}
\end{table}

The proposed incentive-based blockchain service composition technique with reliance on miners, referred to herein as \textit{Incentive-BC1}, is compared against \textit{i)} the same solution without the use of miners, namely, with reliance on participant capability advertisements, referred to herein as \textit{Incentive-BC2}, \textit{ii)} a non-incentive-based BC technique, referred to as \textit{non-Incentive-BC}, and \textit{iii)} a traditional fog-based service composition solution that does not rely on end-devices for service tasks, referred to as \textit{non-BC}. Evaluations were conducted in regards to resource usage, Blockchain formation hit ratio, the delay incurred for the blockchain formation process, and the total amount of rewards shared among participants in the service provisioning process.

\subsection{Resource Usage}
\begin{figure}[h]
	\centering
	\includegraphics[scale=0.35]{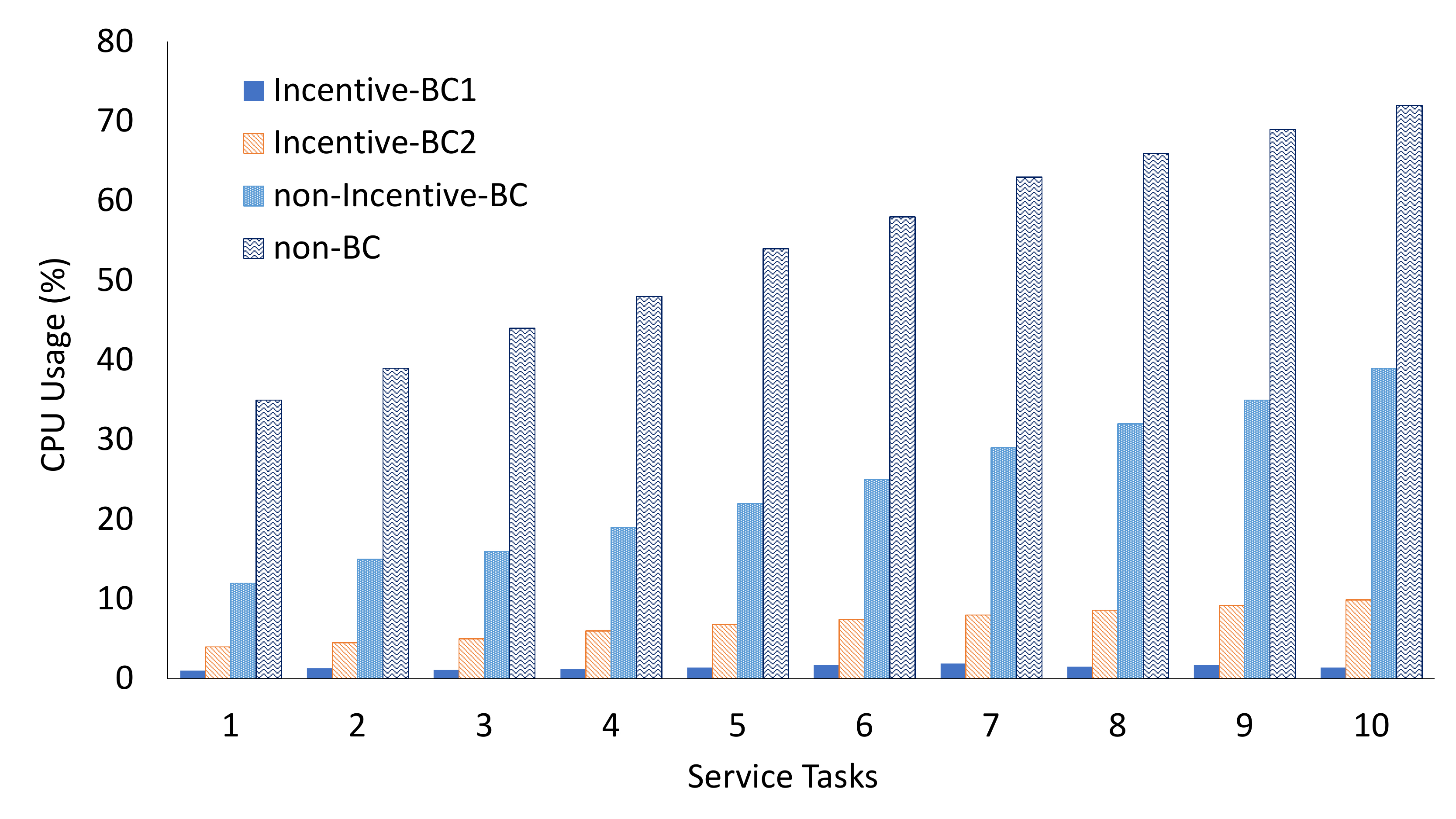}
	\caption{A comparison of the overall CPU usage among four different methods, in terms of reliance on incentives, blockchain and service miners.}
	\label{fig:cpu_usage}
\end{figure}

The amount of resource consumption was based on the average CPU usage per participant (either end-device or fog device) to compose and deliver the requested services. Results depicted in Figure \ref{fig:cpu_usage} show that by relying on the proposed \textit{Incentive-B1} method, the CPU usage per participant sharply drops by more than 70\% when compared against the \textit{non-BC} method. Such a result is very promising and shows that MEC solutions can heavily rely on end-devices to perform service tasks and focus its responsibility on management rather than provisioning. Such a technique will also free up MEC devices to accept more service requests from clients. The use of trusted entities (i.e. miners) provides even further resource enhancements as shown in the figure when comparing the two solutions, namely, \textit{Incentive-BC1} and \textit{Incentive-BC2}. Comparing the two solutions, a reduction of nearly 8\% in CPU resource usage is seen for service requests with 10 service tasks.

\subsection{Energy Consumption}
\begin{figure}[h]
	\centering
	\includegraphics[scale=0.35]{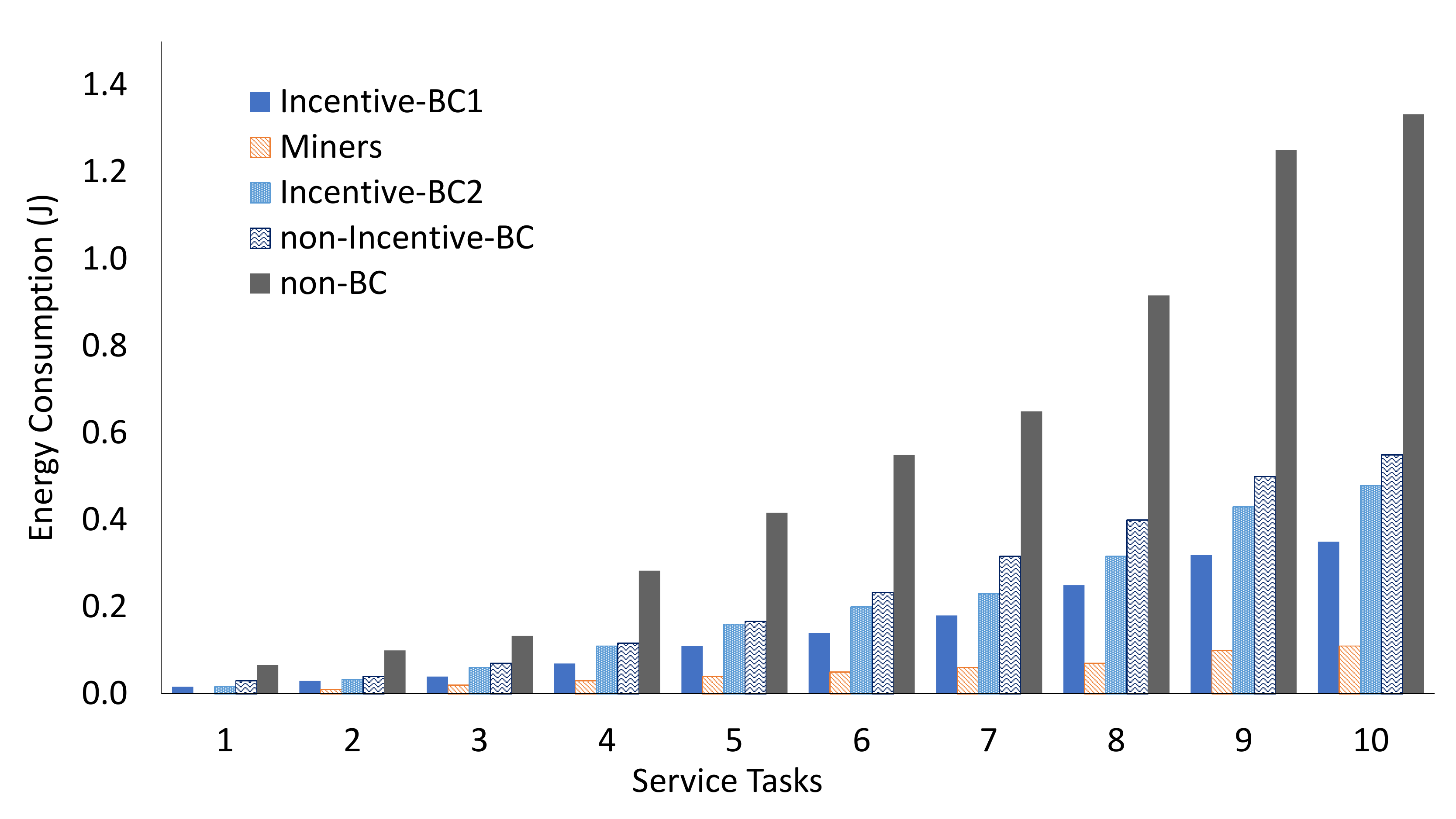}
	\caption{A comparison of the overall energy consumption among four different methods, in terms of reliance on incentives, blockchain and service miners.}
	\label{fig:power_usage}
\end{figure}

An evaluation of the energy consumption of the proposed scheme against the other methods was conducted. The results depicted in Figure \ref{fig:power_usage} are for a network density of 500 end-devices. For the non-BC solution, we assume that all service tasks are available at the fog devices. For the BC solutions, the service tasks are distributed among the end-devices, and hence require node cooperation. Results show that the \textit{Incentive-BC1} (with the aid of miners) and the \textit{Incentive-BC2} both provide similar power consumptions, which outperform the non-incentive mechanisms. It should be noted that although from the figure we see that less power is consumed for the \textit{Incentive-BC1} technique in comparison to \textit{Incentive-BC2}, this reduction is due to the offloaded tasks to miners to select participants in the composition process. The overall reduction of the incentive mechanisms over the non-incentive mechanisms is an overall reduction of nearly 10\% and 140\% in power usage when compared against the \textit{non-Incentive-BC} and \textit{non-BC}, respectively. The incentivized BC solutions have shown that energy consumption at edge nodes is reduced dramatically and is shifted to the end-devices with less energy consumption at the end-device side.

\subsection{Blockchain Formation Hit Ratio}
\begin{figure}[h]
	\centering
	\includegraphics[scale=0.35]{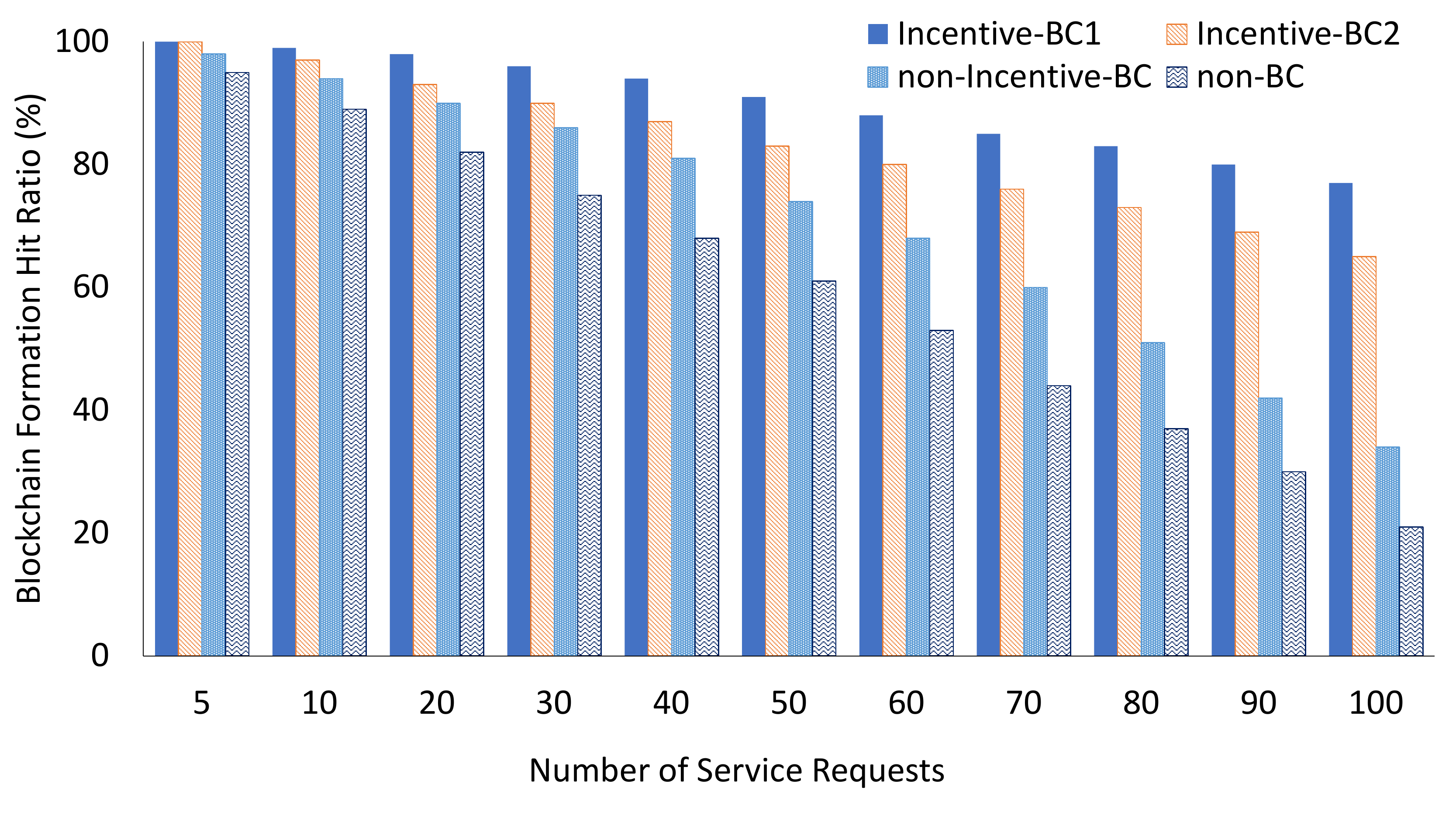}
	\caption{Blockchain formation success rate in terms of the number of simultaneous service requests for four different solutions.}
	\label{fig:hitRatio}
\end{figure}

Testing the effectiveness of the proposed technique in terms of service composition success rate, namely, forming a successful and complete blockchain was considered in one of the experiments. The goal was to increase the number of service requests that arrive simultaneously at the fog devices and observe whether the proposed solution, namely, \textit{Incentive-BC1} can handle excessive amounts of requests. As depicted in Figure \ref{fig:hitRatio}, for the proposed incentive-based solution, the hit ratio is nearly perfect for low to moderate simultaneous service requests. Moreover, for excessive numbers of service requests, precisely with 100 simultaneous service requests, the \textit{Incentive-BC1} solution outperforms all other techniques with nearly 80\% success rate in blockchain formations. Additionally, we see that for the incentive based mechanisms (either with or without miners), the blockchain formation hit ratio is nearly double that of non-incentive techniques, namely, \textit{non-Incentive-BC} and \textit{non-BC}. The hit ratios with 100 simultaneous service requests for \textit{Incentive-BC2}, \textit{non-Incentive-BC} and \textit{non-BC} are 65\%, 34\% and 21\%, respectively.

\subsection{Blockchain Formation Delay}
\begin{figure}[h]
	\centering
	\includegraphics[scale=0.35]{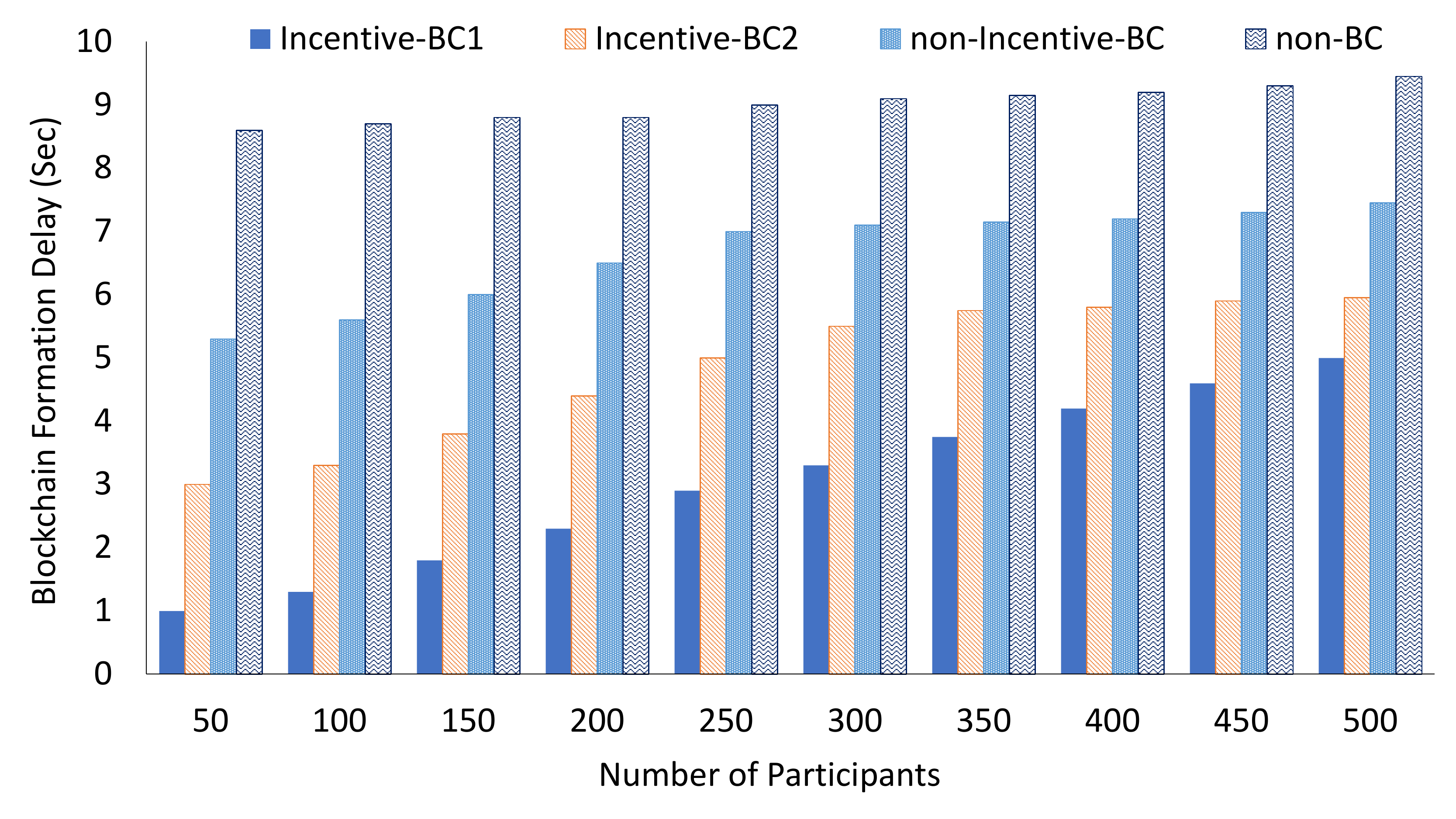}
	\caption{Comparing the delay encountered in forming blockchains as the number of participants varies for four different solutions.}
	\label{fig:delay}
\end{figure}

An experiment was also conducted to determine the delay encountered in forming blockchains (i.e. composing services) as the number of participants vary. A comparison of the proposed solution against the three other techniques is shown in Figure \ref{fig:delay}. The figure shows the average experienced delay, from the initiation of the service request and the completion of the blockchain formation, namely, composition of the requested service. The \textit{Incentive-BC1} solution outperforms the other techniques due to its capability of adapting to the time-constraints of the requested service. Nodes having time-sensitive delay requirements are carried out either at the fog site (if services are available) or carried out by the end-devices with the aid of fog devices and miners. With 500 participants, the blockchain formation delay is reduced by nearly 19\%, 49\%, and 89\% when comparing \textit{Incentive-BC1} against \textit{Incentive-BC2}, \textit{non-Incentive-BC} and \textit{non-BC}, respectively. We note here that some services cannot be composed (i.e. cannot form some blockchains) and hence are not considered in the results. Results in regards to the ratio of non-successful blockchain formations are presented earlier in Figure \ref{fig:delay}.

\subsection{Reward Distribution}
\begin{figure}[h]
	\centering
	\includegraphics[scale=0.35]{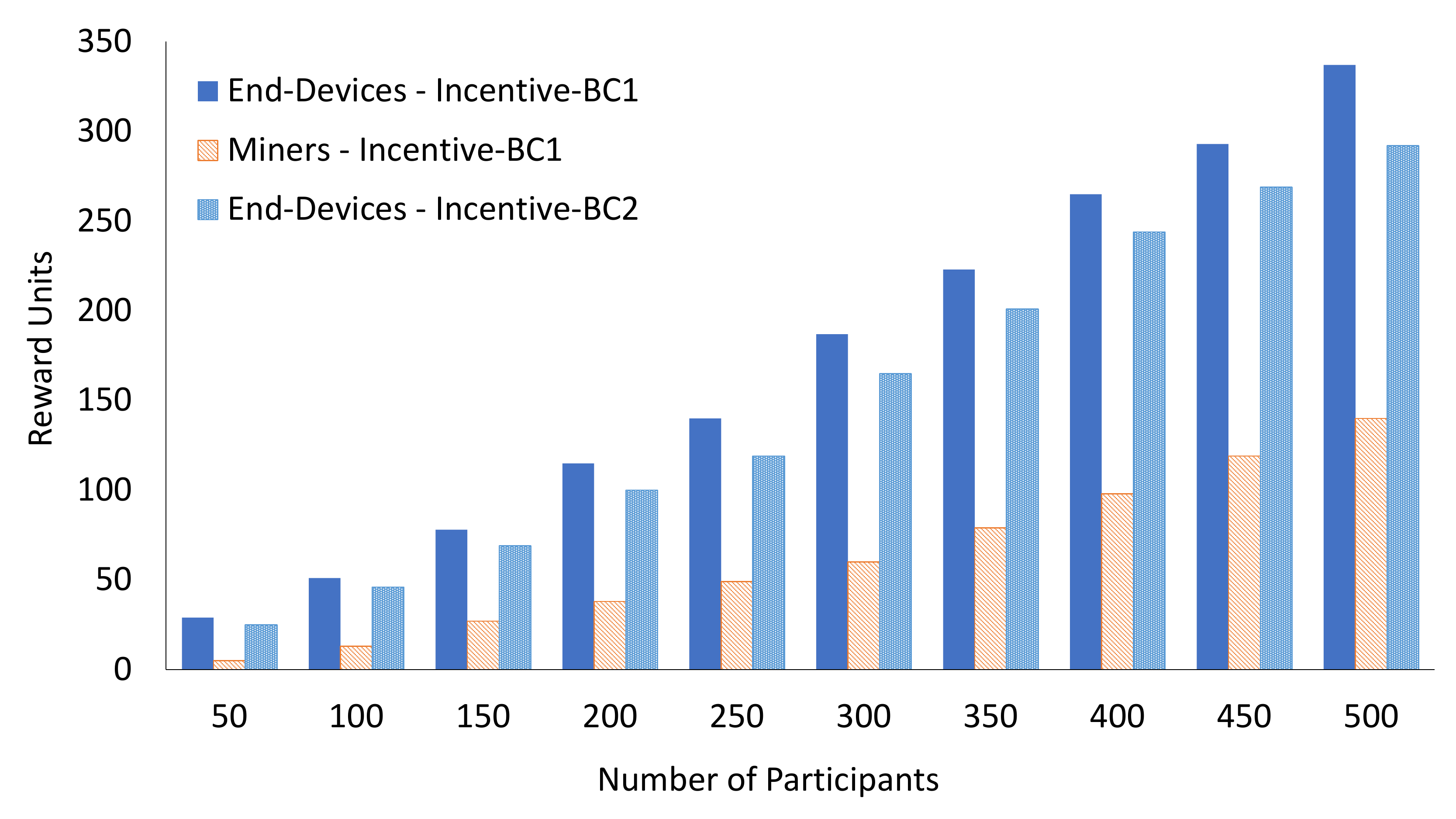}
	\caption{Accumulated rewards for both end-devices and miners using the \textit{Incentive-BC1} and \textit{Incentive-BC2} solutions.}
	\label{fig:reward}
\end{figure}

Reward analysis was conducted on the proposed \textit{Incentive-BC1} solution against the \textit{Incentive-BC2} solution. The main objective of this test is to determine the proportion of rewards gained by miners against end-devices, and whether miners reduce the overall rewards distributed among end-devices. It was evident from the results, depicted in Figure \ref{fig:reward}, that the use of miners in \textit{Incentive-BC1} solution resulted in the accumulation of more rewards for end-devices. This was evident from the increased number of service requests being fulfilled given the aid of miners. For instance, with a simulation run of 500 participants, the total amount of rewards accumulated using the \textit{Incentive-BC1} solution was 477 reward units (337 for end-device and 140 for miners). On the contrary, the total amount of rewards accumulated using the \textit{Incentive-BC2} solution was 292 reward units. That is an increase of 45 reward units just for the end-devices, in addition to the rewards distributed to the miners. By comparing the proportion of rewards distributed among end-devices and miners using the proposed \textit{Incentive-BC1} solution, we see that miners have nearly a third of the rewards in comparison to end-devices.

\section{Conclusion and Future Work}
\label{conclusion}
The vision of beyond 5G communication technologies is to provide connectivity for intelligent connected things. End-devices will have the capability of performing sophisticated intelligent tasks with minimal or in most cases without reliance on centralized or semi-centralized entities like fog, MEC, and cloud computing. Such tactile internet infrastructures will enable seamless and smart interaction between humans and machines to achieve revolutionized solutions for different ecosystems. This article introduced a cooperative IoT framework that relies on blockchain-enabled resource sharing and service composition through volunteer computing. Device capabilities are advertised and made available for sharing using blockchain. Incentives in the form of rewards are are given to participants to ensure fair and balanced cooperative resource usage. Miners are used to search for non-advertised service capabilities to ensure a fast and reliable service provisioning framework. Experimental evaluations conducted in the form of simulations showed that the proposed solution provides adequate and fair distributed rewards to all participants in the blockchain formation process. Moreover, high values of service hit ratio and balanced resource usage among participants was also experienced under the premise of high IoT device availability. For future work, we plan to integrate the concept of federated learning with volunteer computing. IoT devices will collaborate together in the learning process and share their learnt models using blockchains without reliance on any centralized training. This will ensure both data privacy and network security.

\bibliographystyle{unsrt}

\bibliography{acmart}

\end{document}